\documentclass[%
reprint,
 amsmath,amssymb,
 aps,
]{revtex4-1}

\usepackage{graphicx}
\usepackage{dcolumn}
\usepackage{bm}
\usepackage{csquotes}
\usepackage{subfigure}
\usepackage{inputenc}
\usepackage{float}
\begin{document}


\title{Nonlinear Wave Chaos: Statistics of Second Harmonic Fields}

\author{Min Zhou}
\author{Edward Ott}%
\author{Thomas M. Antonsen Jr.}%
\author{Steven M. Anlage}%
\affiliation{%
 Department of Electrical and Computer Engineering, and Department of Physics, University of Maryland, College Park, Maryland 20742-4111, USA
}%

\date{\today}

\begin{abstract}
Concepts from the field of wave chaos have been shown to successfully predict the statistical properties of linear electromagnetic fields in electrically large enclosures. The Random Coupling Model (RCM) describes these properties by incorporating both universal features described by Random Matrix Theory and the system-specific features of particular system realizations. In an effort to extend this approach to the nonlinear domain, we add an active nonlinear frequency-doubling circuit to an otherwise linear wave chaotic system, and we measure the statistical properties of the resulting second harmonic fields. We develop an RCM-based model of this system as two linear chaotic cavities coupled by means of a nonlinear transfer function.  The harmonic field strengths are predicted to be the product of two statistical quantities and the nonlinearity characteristics. Statistical results from measurement-based calculation, RCM-based simulation, and direct experimental measurements are compared and show good agreement over many decades of power.
\end{abstract}

\maketitle
\textbf{The wave properties of systems that show chaos in the classical limit are extensively studied in the field of \enquote{wave chaos}. These systems are quite common in life, ranging from the small scale of atomic nuclei, to mesoscopic quantum systems like quantum dots, to macroscopic acoustic, microwave, optical systems, and even to ocean waves in the sea. The Random Coupling Model (RCM) is a well-established theory that describes the statistical wave scattering properties of these wave chaotic systems. A two-dimensional 1/4-bowtie microwave billiard is one example of a wave chaotic system that has been studied experimentally and numerically. Microwave experiments in these billiards have demonstrated the accuracy of the RCM. However, prior work has been confined to linear systems, and have not taken into account nonlinearity in the wave properties, which is potentially important. By adding an active circuit that generates second harmonics to the 1/4 bowtie-billiard, this work is the first step in trying to generalize the RCM to nonlinear systems, specifically by focusing on the second harmonic field statistics. An extended RCM model that including the characteristics of the nonlinear element shows very good agreement with the experimental results.}
\section{Introduction}
The scattering of short-wavelength waves in domains in which the corresponding rays are chaotic (known as wave chaos) has inspired research activities in many diverse contexts including quantum dots \cite{QD1,QD2}, atomic nuclei \cite{atom}, optical cavities\cite{optic}, microwave cavities \cite{mw1,mw2,mw3}, acoustic resonators \cite{acR1,acR2}, and others. In this case the response is extremely sensitive to the domain's configuration, the driving frequency, and ambient conditions such as temperature and pressure \cite{bini}. Numerical solution of the detailed response of a particular system is computationally intensive and does not necessarily provide much insight to other systems which are slightly different. This leads to the adoption of a statistical description. 

In the field of \enquote{wave chaos}, Random Matrix Theory (RMT) has been shown to successfully describe many statistical properties of bounded wave-chaotic systems (e.g., enclosures such as electromagnetic cavities), including their eigenvalue spectra, eigenfunctions, scattering matrices, delay times, etc. \cite{SH-S,1,2,3,Dietz}. Wave systems also have system-specific features that modify the underlying universal fluctuations. The Random Coupling Model (RCM) accounts for those non-universal features such as the details of ports coupling waves into and out of the domain of the cavity, short orbits that exist between the ports, and specific persistent features of the enclosure \cite{James-shortray, yeh-nonuni,yeh-shortray}. Experimentally, the system-specific features are captured by the impedance (reaction matrix) \cite{Rmatrix} averaged over an ensemble of realizations. By applying this technique to remove non-universal properties, RMT statistical properties have been uncovered in experimental data on ray chaotic 1D quantum graphs \cite{graph}, 2D electromagnetic cavities (known as billiards) \cite{SH-Z} and 3D cavities (reverberation chambers) \cite{Zach-3D}. 

Based on the success of the RCM, it is of interest to explore directions extending its generality. Along this line, theories have been developed for \enquote{mixed} systems which include both regular and chaotic ray dynamics \cite{mix}, and for networks of coupled cavities in which waves propagate from one sub-system to another \cite{Gab-weak,Gab-chain}. While such previous extensions have focused on linear systems, it is of great interest to see how nonlinearity would modify the RCM. The present paper is meant to serve as a first step towards a nonlinear generalization of the RCM by examining the statistics of nonlinearly generated harmonic signals.

Nonlinearity has arisen before in the study of wave chaos. For example, rouge waves can appear in wave chaotic scattering systems \cite{sea1, sea2}. They appear in different physical contexts and are enhanced by nonlinear mechanisms \cite{rogue1, rogue2}.
In acoustics, Time-Reversed Nonlinear Elastic Wave Spectroscopy (TR/NEWS) is based on the nonlinear time reversal properties of a wave chaotic system \cite{8}. TR/NEWS is proposed as a tool to detect micro-scale damage features (e.g., delaminations, micro-cracks or weak adhesive bonds) via their nonlinear acoustic signatures \cite{9,10}. Applying this idea to electromagnetic waves \cite{Fink}, the nonlinear electromagnetic time-reversal mirror shows promise for novel applications such as exclusive communication and wireless power transfer \cite{MF-PRL,MF-PRE}. Theoretical study of stationary scattering from quantum graphs has been generalized to the nonlinear domain, where the nonlinearity creates multi-stability and hysteresis \cite{NLgraph}. A wave-chaotic microwave cavity with a nonlinear circuit feedback loop demonstrated subwavelength position sensing for a perturber inside a cavity \cite{NL2D}. Furthermore, investigation of the electromagnetic field statistics created by nonlinear electronics inside a wave chaotic reverberation chamber has a number of applications, including electromagnetic immunity testing of digital electronics \cite{14,1NL}.

In this work, an active nonlinear circuit is added to a ray-chaotic microwave billiard. The billiard, shown schematically in Fig. \ref{fig1}, has an area of 0.115 $m^2$, and for the microwave wavelengths used here ($3 - 9$ cm) can be considered in the semiclassical or short-wavelength limit. The cavity has a height of $d=7.9$ mm. Thus, below a frequency $f_{max}=c/(2d)=18.9$ GHz, it is a quasi-2D billiard in which the electric field is polarized in the short direction, and the magnetic field is in the 2D plane of the cavity. For frequencies $2f  > 7$ GHz, the mode number is above $\sim200$ and it can be considered that the cavity is in the highly over-moded regime \cite{paulso}. The nonlinear circuit generates and amplifies second harmonic output which is fed back into the billiard. We study the statistics of the second harmonic fields in the cavity for a fixed power at the input fundamental frequency.

\section{Background and Experimental Setup}
In the case of a linear ray-chaotic cavity with N ports, the Random Coupling Model characterizes the fluctuations in the impedance $\overline{\overline{Z}}$  and scattering $\overline{\overline{S}}$  matrices. The cavity impedance $\overline{\overline{Z}}_{cav}$ is related to the average impedance $\overline{\overline{Z}}_{avg}$ through a random variable $\overline{\overline{\xi}}$: 
\begin{equation}
	\overline{\overline{Z}}_{cav}=i\cdot Im(\overline{\overline{Z}}_{avg})+[Re(\overline{\overline{Z}}_{avg})]^{\frac{1}{2}}\cdot\overline{\overline{\xi}}\cdot[Re(\overline{\overline{Z}}_{avg})]^{\frac{1}{2}}
	\label{eq1}
\end{equation}
where $\overline{\overline{Z}}_{avg}$ is an average of impedance over an ensemble of cavity realizations and frequencies. $\overline{\overline{Z}}_{avg}$ contains the system specific features including the radiation impedance of the ports and short orbits that survive the ensemble averages \cite{James-shortray, yeh-nonuni,yeh-shortray}. By inverting Eq. (\ref{eq1}) and subtracting the non-universal features from $\overline{\overline{Z}}_{cav}$ in each realization, one can uncover $\overline{\overline{\xi}}$ which are universal fluctuations described by Random Matrix Theory (RMT). The loss parameter $\alpha$ is the only parameter determining the statistics of the universal fluctuations. $\alpha={k}^{2}A/(4\pi Q)$ is the ratio of the mean 3-dB bandwidth of the resonant modes to the mean spacing in frequency between the modes. $k=2\pi f/c$ is the wave number of frequency $f$, $A$ represents the area of the billiard, and $Q$ is the typical loaded quality factor of the enclosure with the assumption that losses are uniform. Experimental tests in various wave chaotic systems have systematically explored the effects of different loss parameters on the statistical properties, ranging from cryogenic superconducting cavities ($\alpha \sim 0.01)$ \cite{yeh-insitu,yeh-lowloss} to three-dimensional complex enclosures ($\alpha > 10$)  \cite{Zach-3D}.
\begin{figure}
  \begin{center}    
    \subfigure{\label{fig3a}\includegraphics[width=0.7\linewidth]{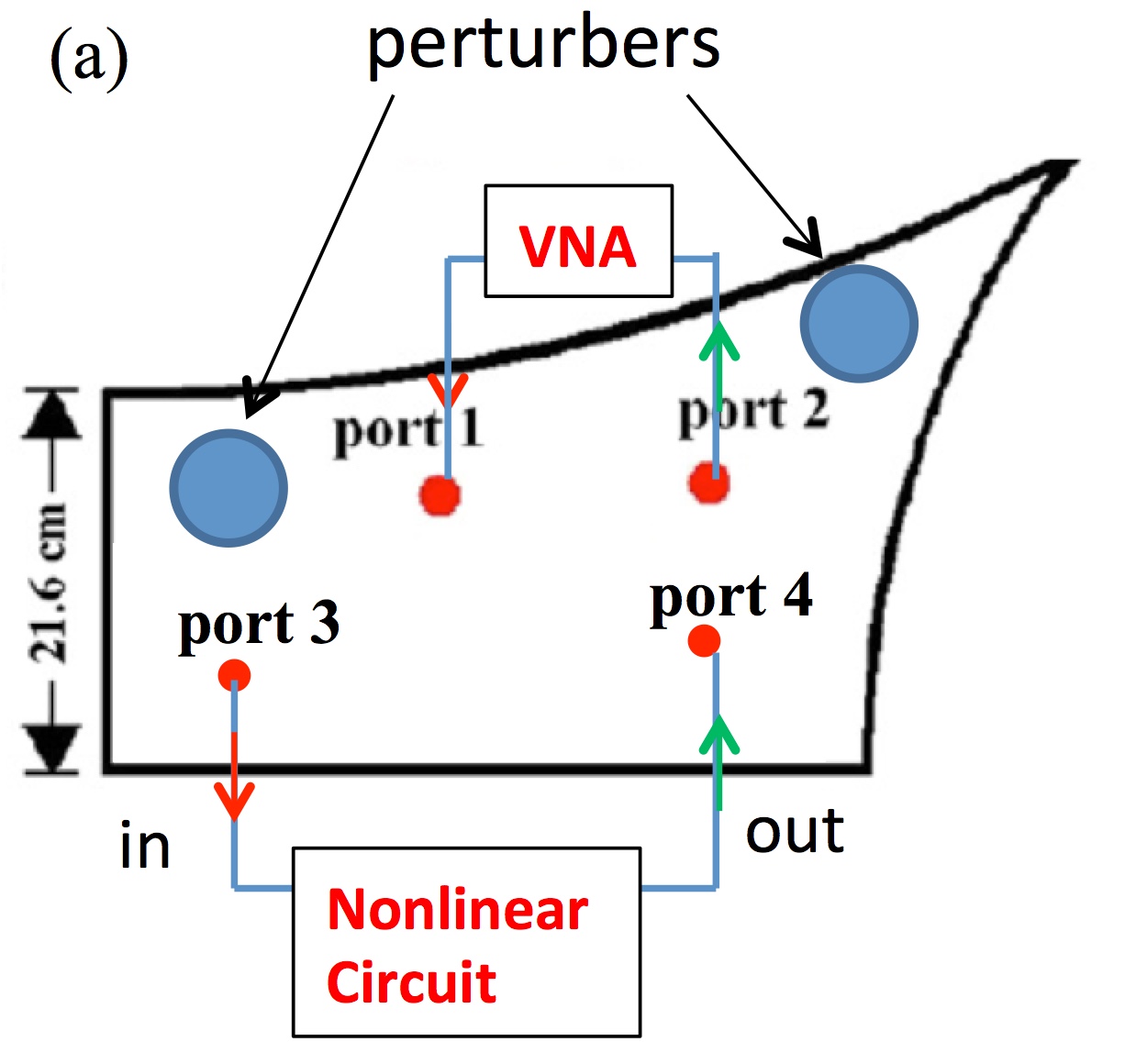}} 
    \subfigure{\label{fig3b}\includegraphics[width=\linewidth]{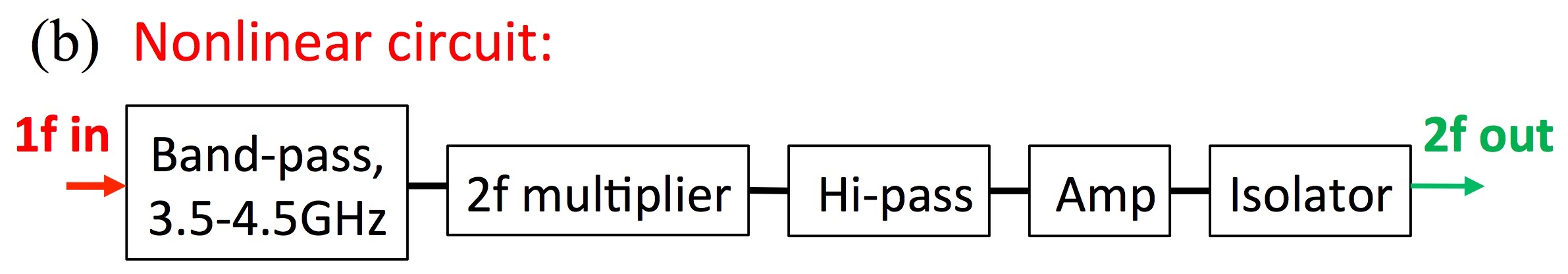}}
  \end{center}
	\caption{Experimental setup: $1/4$-bowtie cavity with an active nonlinear circuit. (a) The vector network analyzer (Model: Keysight PNA N5242A or E8364C) measures the absolute power of harmonics at port 2 relative to the input fundamental tone at port 1. (b) The active nonlinear circuit consists of two cascaded 3.5 GHz - 4.5 GHz band pass filters (Mini-Circuit VBFZ-4000-S+), a frequency doubler (Mini-Circuit ZX9C-2-50-S+), two cascaded high pass filters (mini-circuit VHF-6010+), a wide band power amplifier (HP83020A) and an isolator (FairviewMicrowave SFI 0418).}
  \label{fig1}
\end{figure}

Here a symmetry-reduced \enquote{$1/4$-bowtie} shape (Fig. \ref{fig1}) quasi-two-dimensional cavity at room temperature is used as the ray chaotic system \cite{paulso}. To introduce nonlinearity, an active nonlinear circuit is connected to two ports of the billiard as shown in Fig. \ref{fig1}. The active nonlinear circuit is designed to double the input frequency in the range from 3.5 GHz to 4.5 GHz; other harmonics, as well as the fundamental tone, are suppressed at the output. Measurements are taken between two additional ports of the cavity, and an ensemble of billiard realizations is created by moving two perturbers throughout the cavity. Thus the realizations maintain a fixed volume and mean mode spacing. A sinusoidal tone at fundamental frequency $1f$ with a certain power is created in the Vector Network Analyzer (VNA) and injected through port 1. Port 3 is the input of the active nonlinear circuit. Due to the ray-chaotic properties of the cavity, the $1f$ signal received by port 3 varies over several decades in power as a function of frequency and perturber locations. The output at port 4 will be at the 2nd harmonic frequency with a certain power. Port 4 serves as the source of a $2f$ signal injected into the cavity. The VNA, set in Frequency Offset Mode (FOM) picks up the 2nd harmonics and measures the absolute power through port 2. 

\section{Model}
We separately characterize the nonlinear circuit under FOM and find that for input powers in the range -45 dBm to -5 dBm, ${P}{_{out,2f}} \, vs. \, {P}{_{in,1f}}$ obeys a simple empirical relation:
\begin{equation}
	{P}{_{out,2f}}=slope\cdot{P}{_{in,1f}}+intercept
	\label{eq2}
\end{equation}
where $slope = 2.00 \pm 0.01$ and the amplifier contributes to the \enquote{intercept} term. Note that power $P$ is in dBm and the \enquote{intercept} here refers to intercept in units of dBm, i.e. ${P}{_{out,2f}}$ when ${P}{_{in,1f}}$ = 0 dBm. In terms of power measured in watts, this relation is effectively ${P}{_{out,2f}}={P}{_{in,1f}^2}/P_{norm}$.

\begin{figure}
	\includegraphics[width=\linewidth]{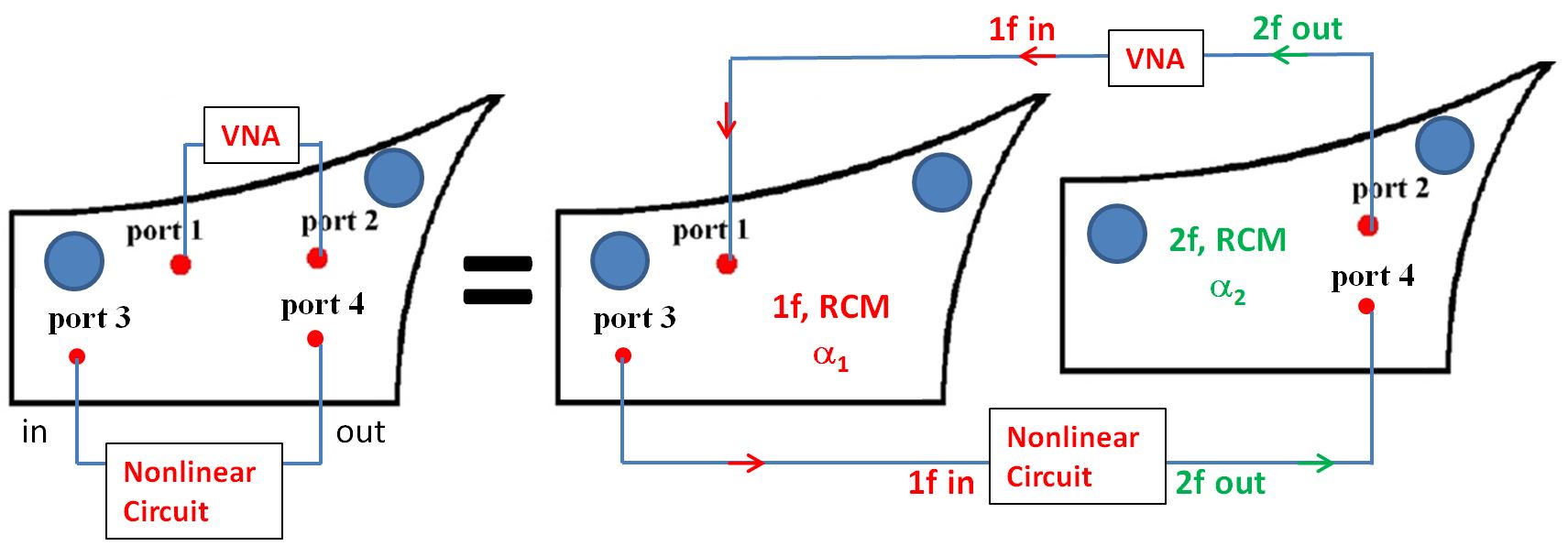}
	\caption{Model of the nonlinear billiard in terms of cascaded cavities. The bowtie with an active nonlinear circuit attached can be considered as two linear billiards operating at different frequencies and loss parameters coupled through the nonlinear circuit}
	\label{fig2}
\end{figure}

To describe the statistical properties of the second harmonic signals measured at port 2, a model of two cascaded linear cavities connected through the nonlinear circuit is proposed (see Fig. \ref{fig2}). This choice was motivated by earlier work on the statistics of impedance and power fluctuations in chains of wave chaotic cavities connected by weak but linear coupling \cite{Gab-weak,Gab-chain,RCMreview}. As shown in Fig. \ref{fig2}, the source signal enters the cavity from port 1. The signal reaching port 3, which is the input of the nonlinear circuit, is given by the linear S-parameters between port 1 and 3, denoted by ${S}_{13}$. Its statistics are described by the linear Random Coupling Model at $1f$ with loss parameter $\alpha_{1}$. The output signals of the nonlinear circuit at port 4 are at the 2nd harmonic of the input at port 3. Their relation is characterized by the empirical law of the active nonlinear circuit, Eq. (\ref{eq2}). Lastly, the 2nd harmonic signals received at port 2 are linearly related to the 2nd harmonic signals introduced at port 4, which is given by the statistical fluctuations of the S-parameters between port 2 and port 4 denoted by ${S}_{24}$. The statistics of $S_{24}$ are described by the linear Random Coupling Model at frequency $2f$ with loss parameter $\alpha_{2}$. Since the vector network analyzer in FOM measures power, we have a simple relation for the power of 2nd harmonics received at port 2,
\begin{equation}
	{P}{_{out,2f}}=slope\cdot({P}{_{in,1f}}+{|{S}_{13}|}^2)+intercept+{|{S}_{24}|}^2
	\label{eq3}
\end{equation}
where ${P}{_{in,1f}}$, $slope$ and $intercept$ are deterministic; and ${|{S}_{13}|}^2$, ${|{S}_{24}|}^2$ are fluctuating quantities measured in dB.

\section{Results}
\begin{figure}
  \begin{center}    
    \subfigure{\label{fig3a}\includegraphics[width=\linewidth]{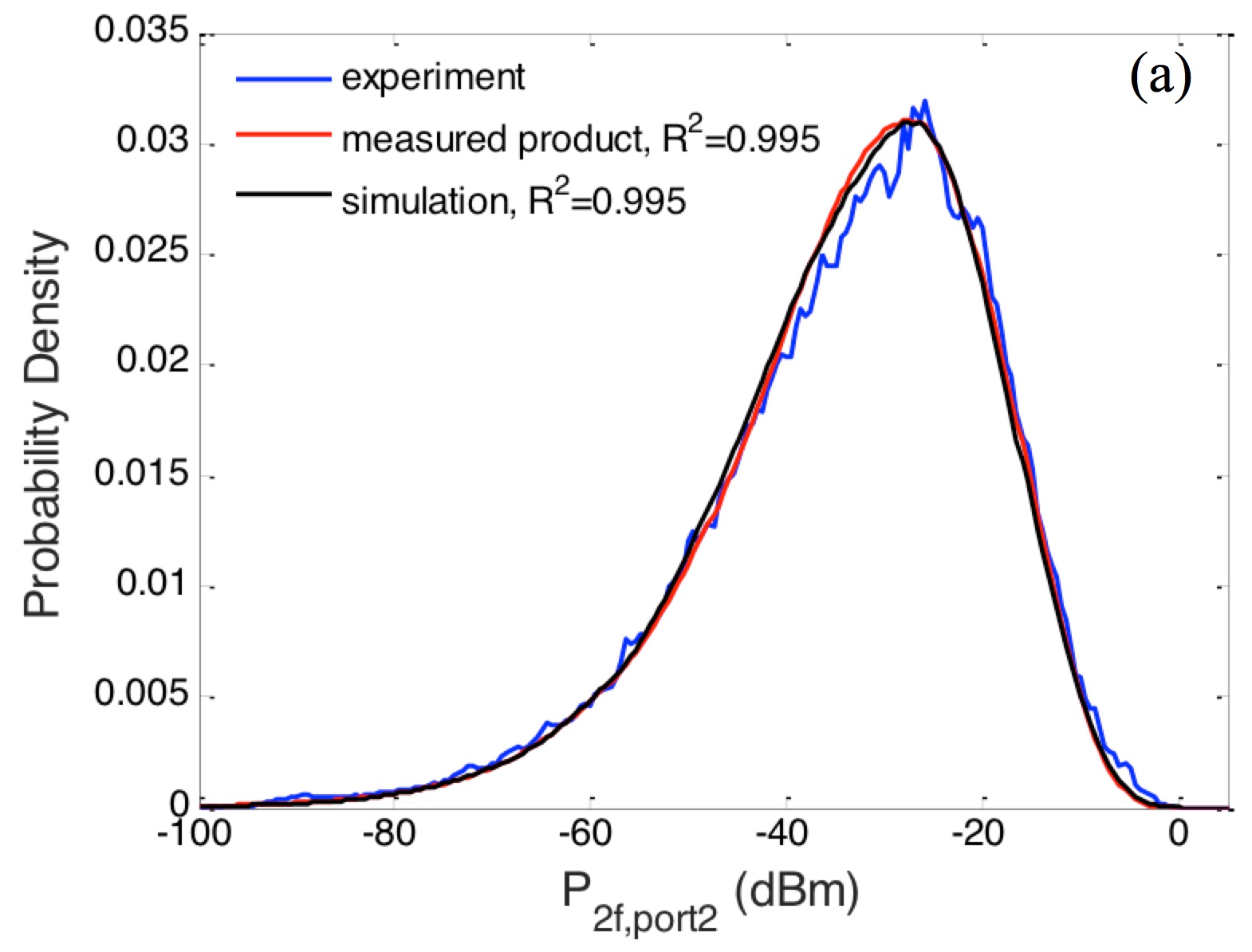}} 
    \subfigure{\label{fig3b}\includegraphics[width=\linewidth]{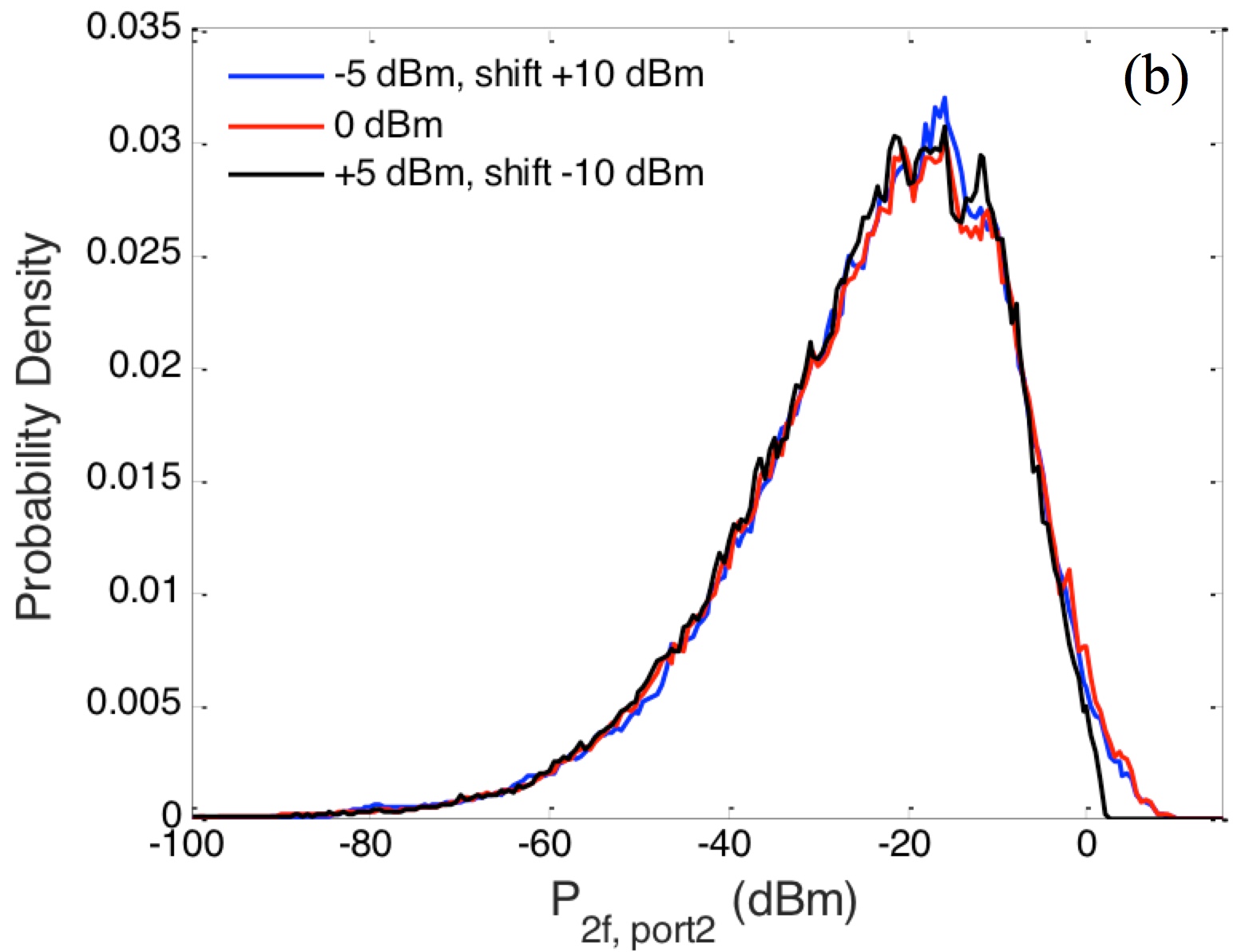}}
  \end{center}
  \caption{Measurement of second harmonic power statistics and test of the model of cascaded cavities. (a) The statistics of the output 2nd harmonic power predicted by the model are compared with direct measurement results (blue). The statistics are compiled over a 2 GHz range with center frequency of 8 GHz. (b) Three input powers are measured: -5 dBm, 0 dBm and +5 dBm, and the distributions are shifted by 10 dBm to overlap.The cutoff near 0 dBm for the +5 dBm curve comes from the saturation of the VNA (Model E8364C). For ${P}{_{in,1f}}=-5$ dBm, 0 dBm and 5 dBm, the mean value is -33.6 dBm, -23.6 dBm and -14.3 dBm, respectively. The standard deviations are 14.6 dBm, 14.8 dBm and 14.4 dBm, respectively.}
  \label{fig3}
\end{figure}

In the results shown below, there are 136 realizations for $P_{in, 1f}=-5$ dBm input, 99 for 0 dBm, and 91 for +5 dBm respectively. The second harmonic signal is measured and histograms are complied over the ensemble of realizations as well as second harmonic frequency between 7 - 9 GHz. These histograms are plotted in Fig. \ref{fig3} and will be compared with theory predictions.

To test the extended RCM model, we have two approaches. The \enquote{measured product} is a calculation based on separate measurement of each linear component, i.e. measurements of ${S}_{13}$, ${S}_{24}$. S-parameters between ports 1 and 3 are measured at 3.5 - 4.5 GHz, for 120 realizations of the positions of the perturbers. S-parameters between ports 2 and 4 are measured at 7 - 9 GHz, again for 120 different realizations of the perturber positions. In the direct experimental measurement of the $2f$ signal with the $1f$ input, the $1f$ and $2f$ signals pass through the cavity with the same perturber position. However, in the \enquote{measured product}, the S-parameters ${S}_{13}$ and ${S}_{24}$ are measured independently, each with a separate ensemble of perturber positions. Their values will not correspond directly to those in the case in which the entire transfer function is characterized. Statistically, the histograms for the \enquote{measured product} and \enquote{experiment} will correspond if ${S}_{13}$ and ${S}_{24}$ are effectively independent.  By putting the measured quantities into the relation Eq. (\ref{eq3}), we create $120^2$ \enquote{realizations} of ${P}{_{out,2f}}$. We call this a \enquote{super data set} and its $P_{2f}$ statistics can be compared with those measured directly.

Another approach, termed \enquote{simulation}, utilizes the RCM to generate a statistical distribution of power values. By using the measured ensembles of ${S}_{13}$ and ${S}_{24}$ mentioned above, the RCM formulation (Eq. (\ref{eq1})) can be applied to extract $Z_{avg}(1f$, port 1 and 3), $Z_{avg}(2f$, port 2 and 4) and loss parameters ${\alpha}_1$, ${\alpha}_2$, respectively. Having the $Z_{avg}$ and loss parameters, we can perform Monte Carlo RMT simulations, to generate ensembles of ${S}_{13}$ and ${S}_{24}$, also using RCM (Eq. (\ref{eq1})). Again, we generate 120 realizations of ${S}_{13}$ and ${S}_{24}$ respectively, and substitute them into Eq.(\ref{eq3}) to create a \enquote{super data set}. The result is based on simulated universal quantities \enquote{dressed} by the measured non-universal features.

The loss parameters $\alpha$ at $1f$ and $2f$ are both less than 1. In such a low loss chaotic environment, the individual cavity mode contributions to the S-parameters will be sharp and distinct. The value of ${P}{_{in,1f}}$ at port 1 is set so that the majority $1f$ power at port 3 falls within the range where Eq. (\ref{eq2}) holds. As a result, the power of the second harmonic signal spans a wide range. Figure \ref{fig3} shows histograms of second harmonic power plotted on a log scale with units of dBm. The histogram is compiled from an ensemble of realizations and over a $2f$ output frequency range from 7 to 9 GHz. For a fixed input power over a certain frequency band, the 2nd harmonic output power varies over 8 - 10 decades. For example, Fig. \ref{fig3a} shows the result for input power ${P}{_{in,1f}}=-5$ dBm. The blue curve is the histogram from direct measurement, where the received power of the 2nd harmonics varies from -100 dBm to 0 dBm. It has a mean of -14.3 dBm and standard deviation of 14.4 dBm. The red curve is the \enquote{measured product}, also derived from measurement. Although derived by a different approach, the overall statistics agrees very well with our direct measurement of the second harmonic power. The black curve labeled \enquote{simulation} is created based on the RCM as described above, and as shown in the plots, it agrees quite well with the red and blue curves, demonstrating the validity of the Random Coupling Model. To quantify the agreement, $R^2$ with respect to the experimental curve is calculated for each model histogram. The red curve \enquote{measured product} has $R^2=0.995$ and the black curve \enquote{simulation} has $R^2=0.994$, both indicating very good agreement. We emphasize that this model yielding this agreement has \textit{no fitting parameter}.

The model (Eq. (\ref{eq3})) predicts that changing the input power should simply shift the PDFs of ${P}{_{out,2f}}$ by 10 dB for each 5 dB increase in input power. Fig. \ref{fig3b} shows the shifted curves of experimental results with respect to the 0 dBm case. Indeed the overall distribution has a similar shape for each input power. However, experimentally the VNA reaches its maximum detectable power at nearly 15 dBm. This is why there is a cutoff at high power for the curve of ${P}{_{in,1f}}=5$ dBm. Results of RCM-based model fits to the second harmonic statistics for input power $P_{in,1f}$ = 0 dBm and 5 dBm are shown in the supplemental material \cite{sup}.

\section{Further Analysis and Discussion}
Prior work \cite{1NL} has investigated second harmonic generation by nonlinear electronics irradiated in a reverberation chamber. The statistics of the re-radiated harmonic spectrum were investigated by using a model of cascaded random processes. Under the assumption that the distribution of the linear field statics in the reverberation chamber follow a Rayleigh distribution, a combined Rayleigh distribution was derived for the statistics of the harmonics \cite{1NL}. However, it has been shown that the Rayleigh distribution of the field statistics in a cavity is only valid in the high-loss regime (loss parameter $\alpha \gg 1$) \cite{yeh-fading,yeh-1st}. By introducing microwave absorbers to the perimeter of the billiards \cite{SH-Z,SH-uni2} or putting the billiard in a dry ice low temperature environment, we are able to tune the loss parameter $\alpha$ from 0.1 to 6 or higher. We show in the supplemental material that the Rayleigh model for linear S-parameter statistics fails in the low loss environment while RCM succeeds \cite{sup}. As a result, the RCM-based model shows much better agreement with the 2nd harmonic field statistics than the combined Rayleigh model. Further, a more versatile \enquote{Double Weibull} model was developed that takes into account the saturation of the nonlinear element \cite{describe,DW}. The fitting parameter in this model is directly related to the nonlinear exponent ($slope$ in Eq. (\ref{eq2})). Our results show that the fitting parameter changes dramatically with loss in the billiard (for a fixed nonlinear transfer function), and that the values are un-physical. The detailed results are given in the supplemental material \cite{sup}. The RCM-based model holds in both high loss and low loss cases, without any fitting parameters, and thus we conclude that the RCM-based model is a more accurate and physically realistic description of the system.

\section{Conclusions}
In summary, by adding an active nonlinear circuit to the ray-chaotic 1/4-bowtie cavity, and extending the Random Coupling Model, it is possible to predict the statistics of harmonics in a nonlinear wave chaotic system. The model of nonlinear cascaded cavities Eq. (\ref{eq3}), which incorporates nonlinearity into the Random Coupling Model describes the effects of the active nonlinear circuit, and is valid both in the low loss and high loss regimes.

\section{Supplemental Material}
See supplemental material for (I) Results of the extended RCM model (Eq. (\ref{eq3})) fits to second harmonic power statistics of input power $P_{in,1f}$ = 0 dBm and +5 dBm, respectively.  (II) Results of Comparison with other models (Rayleigh model to linear fields, Combined Rayleigh and Double Weibull model to second harmonic fields).

\section{Acknowledgements}
This work is supported by ONR under Grant No. N000141512134, AFOSR under COE Grant FA9550-15-1-0171, COST Action IC1407 \lq{ACCREDIT}\rq{ }supported by COST (European Cooperation in Science and Technology), and the Maryland Center for Nanophysics and Advanced Materials (CNAM).

\end{document}



\title{(Supplemental) Nonlinear Wave Chaos: Statistics of Second Harmonic Fields}

\author{Min Zhou}
\author{Edward Ott}%
\author{Thomas M. Antonsen Jr.}%
\author{Steven M. Anlage}%
\affiliation{%
 Department of Electrical and Computer Engineering, and Department of Physics, University of Maryland, College Park, Maryland 20742-4111, USA
}%

\date{\today}

\maketitle
\renewcommand{\thefigure}{S\arabic{figure}}
\section{Tests of the RCM-based model}
\begin{figure}[H]
  \begin{center}    
    \subfigure{\label{fig1b}\includegraphics[width=0.44\linewidth]{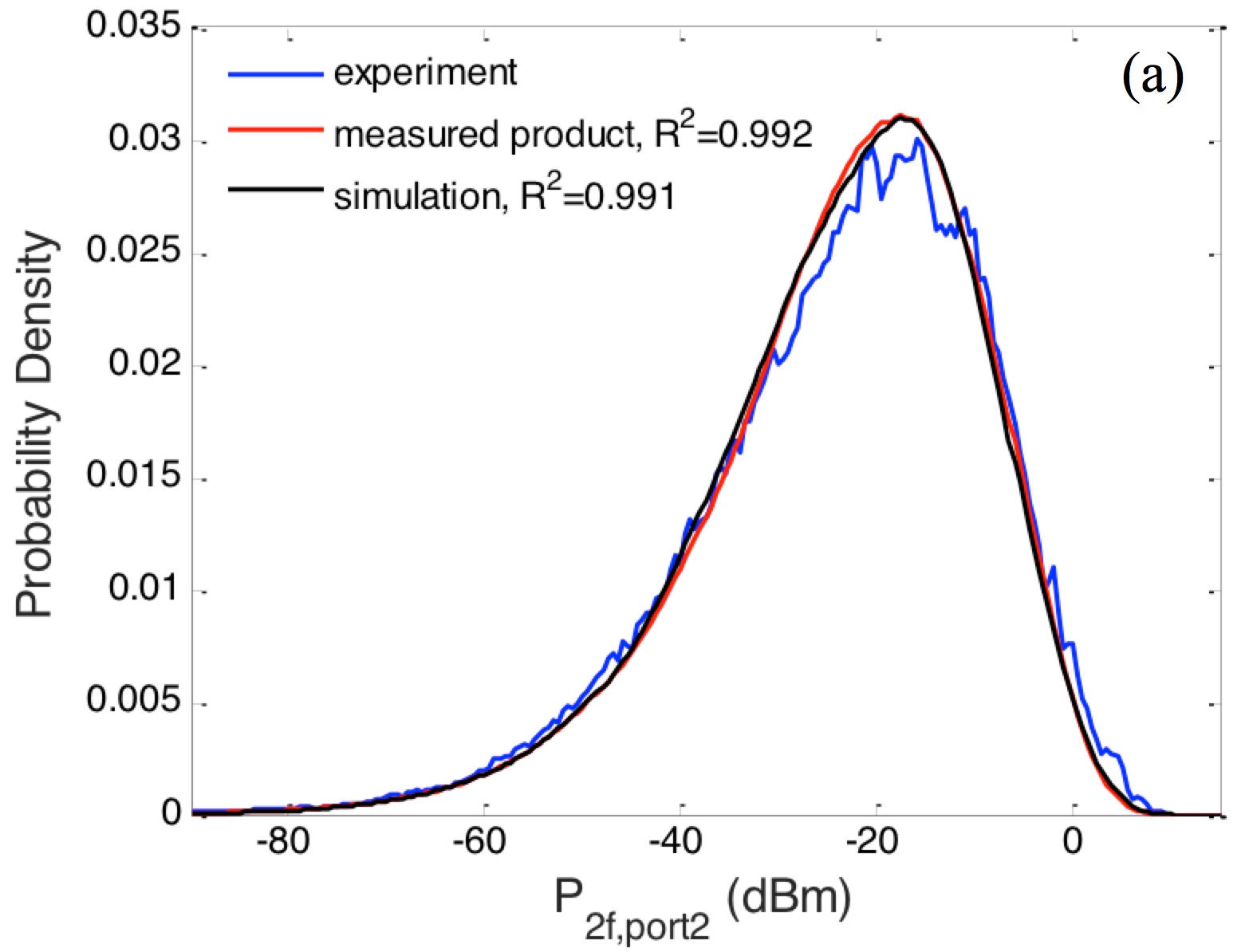}}
    \subfigure{\label{fig1c}\includegraphics[width=0.445\linewidth]{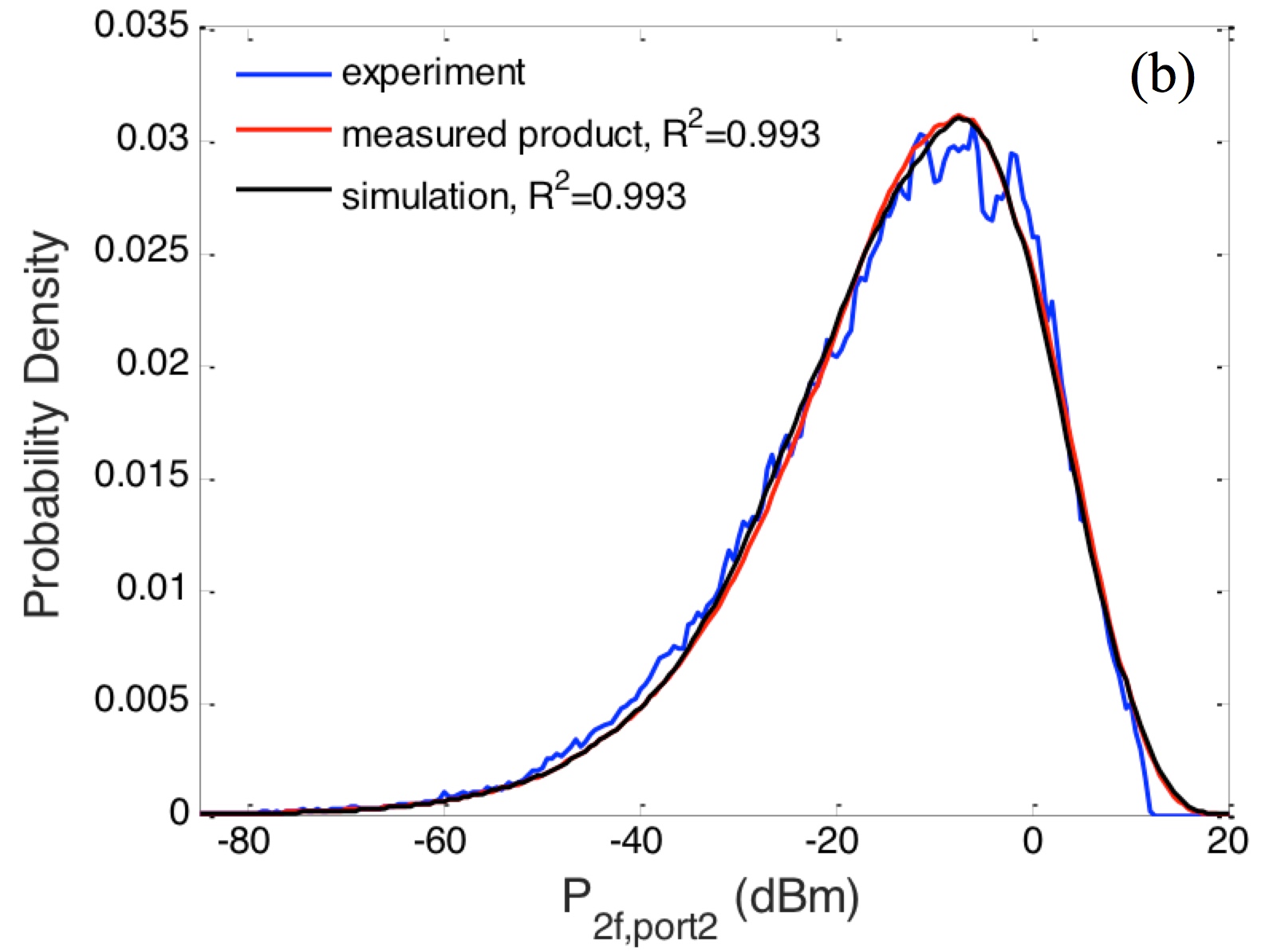}}
  \end{center}
  \caption{Measurement of second harmonic power statistics and test of the model of cascaded cavities. The statistics of the output power of the 2nd harmonics predicted by the model are compared with direct measurement results (blue). The statistics are compiled over a 2 GHz range with center frequency 8 GHz. Two input powers are shown: (a) 0 dBm and (b) +5 dBm.The higher power ($\sim +12$ dBm) cutoff for +5 dBm curve comes from the saturation of the VNA (Model E8364C).}
  \label{fig1}
\end{figure}

Here we show additional data similar to that in Fig. 3 of the main text. Fig. S\ref{fig1b} and S\ref{fig1c} show the results for statistics of second harmonic power for ${P}{_{in,1f}}=0$ dBm and ${P}{_{in,1f}}=+5$ dBm respectively. The measured statistics agree with \enquote{measured product} and \enquote{simulation} models. To quantify the agreement with the data, at ${P}{_{in,1f}}=0$ dBm, the \enquote{measured product} has $R^2=0.992$, and the \enquote{simulation} has $R^2=0.991$. For the ${P}{_{in,1f}}=+5$ dBm data, \enquote{measured product}: $R^2=0.993$, \enquote{simulation}: $R^2=0.993$. 

\section{Comparison with other models}
\subsection{Rayleigh distribution of electric fields}
We first focus on the statistics of linear fields in the 1/4-bowtie cavity. Ref. \cite{1NL} assumes that the distribution of the magnitude of the electric field strength in a reverberation chamber follows a Rayleigh distribution. To test this model in our 1/4-bowtie cavity, the experiment has been conducted in various loss cases, ranging from a dry ice environment ($T \sim -78^{\circ}$C) to reach the low loss regime (loss parameter $\alpha_1=0.31$ at $1f$, $\alpha_2=0.82$ at $2f$) to a high loss environment (loss parameter $\alpha_1=2.9$, $\alpha_2=4.6$ at $2f$) by adding microwave absorber segments along the perimeter of the cavity at room temperature. As loss increased, both the mean and standard deviation of $P_{out,2f}$ were seen to decrease. 
\begin{figure}[H]
  \begin{center}    
    \subfigure{\label{fig5a}\includegraphics[width=0.45\linewidth]{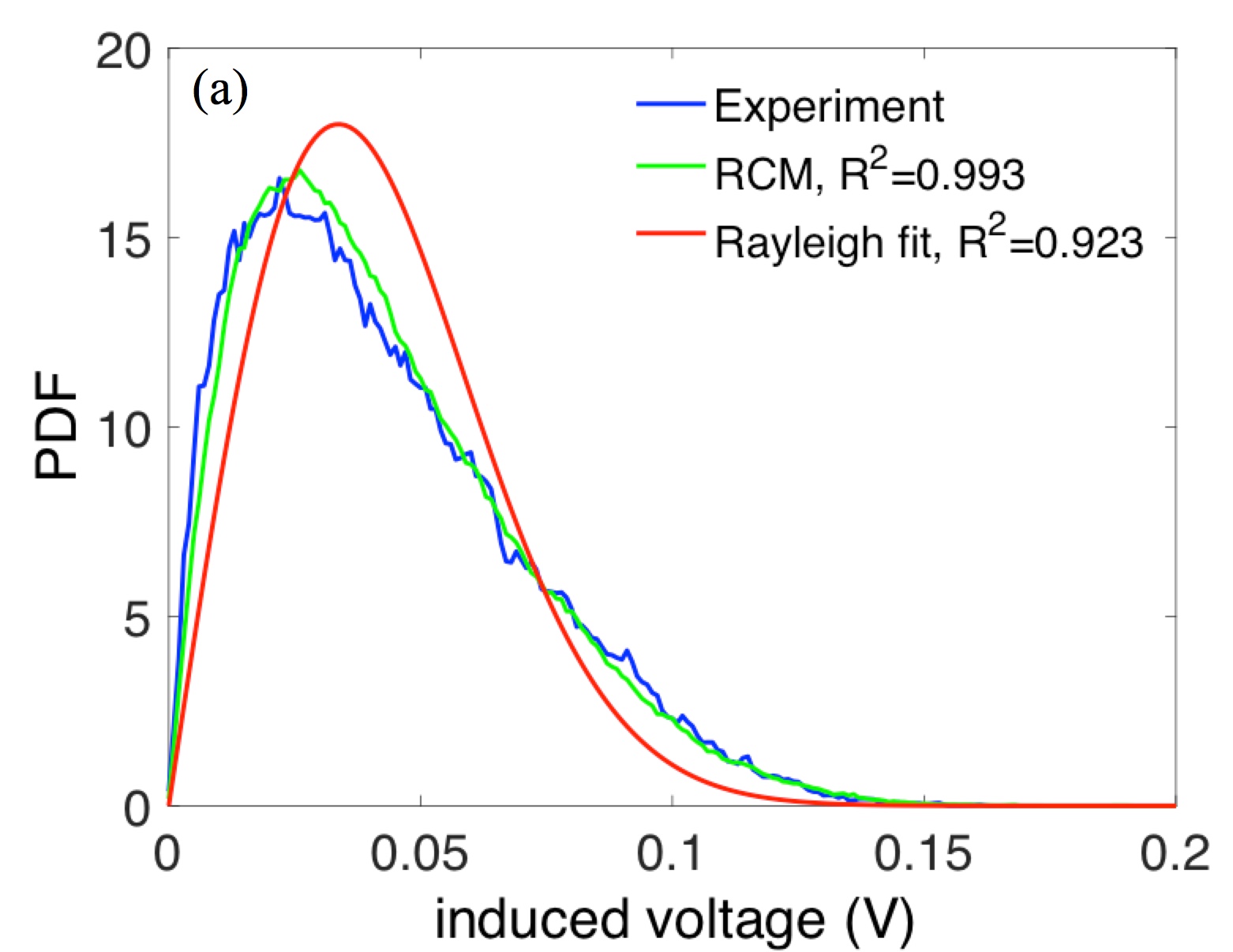}}
    \subfigure{\label{fig5b}\includegraphics[width=0.45\linewidth]{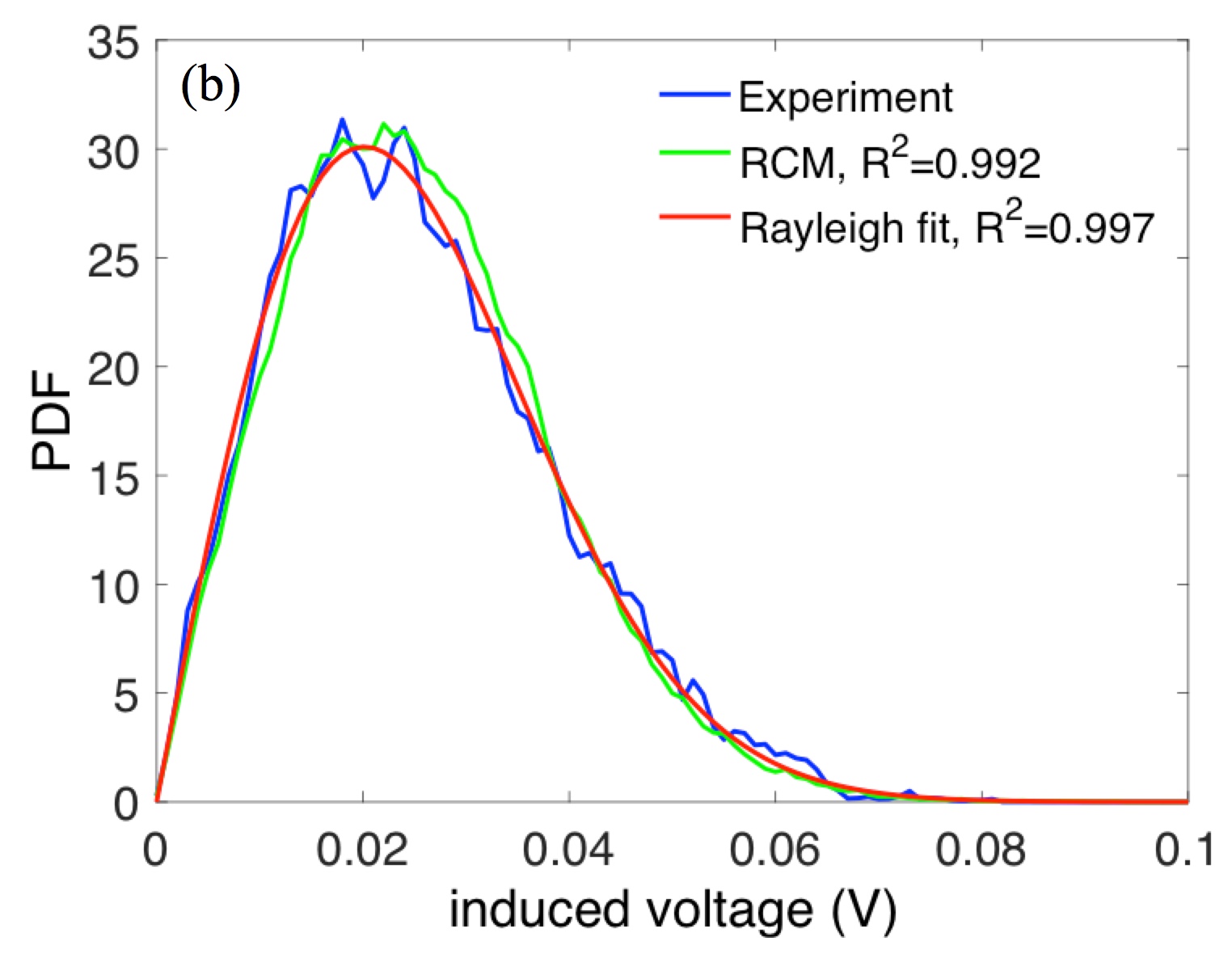}}
  \end{center}
	\caption{Best Rayleigh fit (red) by varying the scale parameter $\sigma$ to the linear induced voltage in the load impedance between port 1 and 3. Singular RCM prediction (green) is shown for comparison. It is generated with known $Z_{avg}$ and loss parameter $\alpha_1$, and by applying Eq. (1) in the main text with Monte Carlo simulation. (a) low loss case $\alpha$ = 0.37,$R^2(RCM) = 0.993$, $R^2(Rayleigh) = 0.923$; and (b) high loss case $\alpha$ = 2.50, $R^2(RCM) = 0.992$, $R^2(Rayleigh) = 0.997$. Frequency range 3.5 GHz - 4.5 GHz.}
	\label{fig5}
\end{figure}

The Rayleigh distribution relies on a scale parameter $\sigma$ (not shown in the figure). Fig. \ref{fig5} shows the best Rayleigh fit (red) to the linear field statistics between port 1 and 3 in low loss (Fig. S\ref{fig5a}) case with loss parameter $\alpha=0.37$ and high loss case (Fig. S\ref{fig5b}) with $\alpha=2.50$. Here the Rayleigh fit is applied to the induced voltage at the receiving port since the induced voltage is a linear function of the electric field. To calculate the induced voltage, the power is converted to voltage by considering a 50 $\Omega$ load impedance at the receiving port. As a comparison, the RCM singular prediction is generated (green). Without additional fitting parameters, the RCM prediction is generated with known $Z_{avg}$ and loss parameter $\alpha_1$. Then Eq. (1) of the main text is applied by performing Monte Carlo RMT simulation. Fig. \ref{fig5} shows that, as expected \cite{yeh-fading,yeh-1st}, the Rayleigh fit deviates from the experiment in the low loss case and fits well to the high loss case, while the RCM prediction agrees well in both cases. In quantifying the agreement, $R^2$ for RCM are both larger than 99\%, while $R^2=0.923$ in low loss and $R^2=0.997$ in high loss for Rayleigh fit. 

\subsection{Combined Rayleigh distribution of second harmonic fields}
We now consider the statistics of second harmonics in the case shown in Fig. 1 of the main text. The Combined Rayleigh distribution \cite{1NL} is based on the assumption that the linear fields obey a Rayleigh distribution. The nonlinear element is modeled as a reradiating source for harmonics, and the generated harmonic signal is related to the incident signal by a power series for weak nonlinearity, i.e. for second harmonic signal, $V_{2f} \propto V_{1f}^2$. The received harmonic signal is therefore the result of a cascaded Rayleigh process. 
\begin{figure}[H]
\begin{center}    
    \subfigure{\label{fig6a}\includegraphics[width=0.45\linewidth]{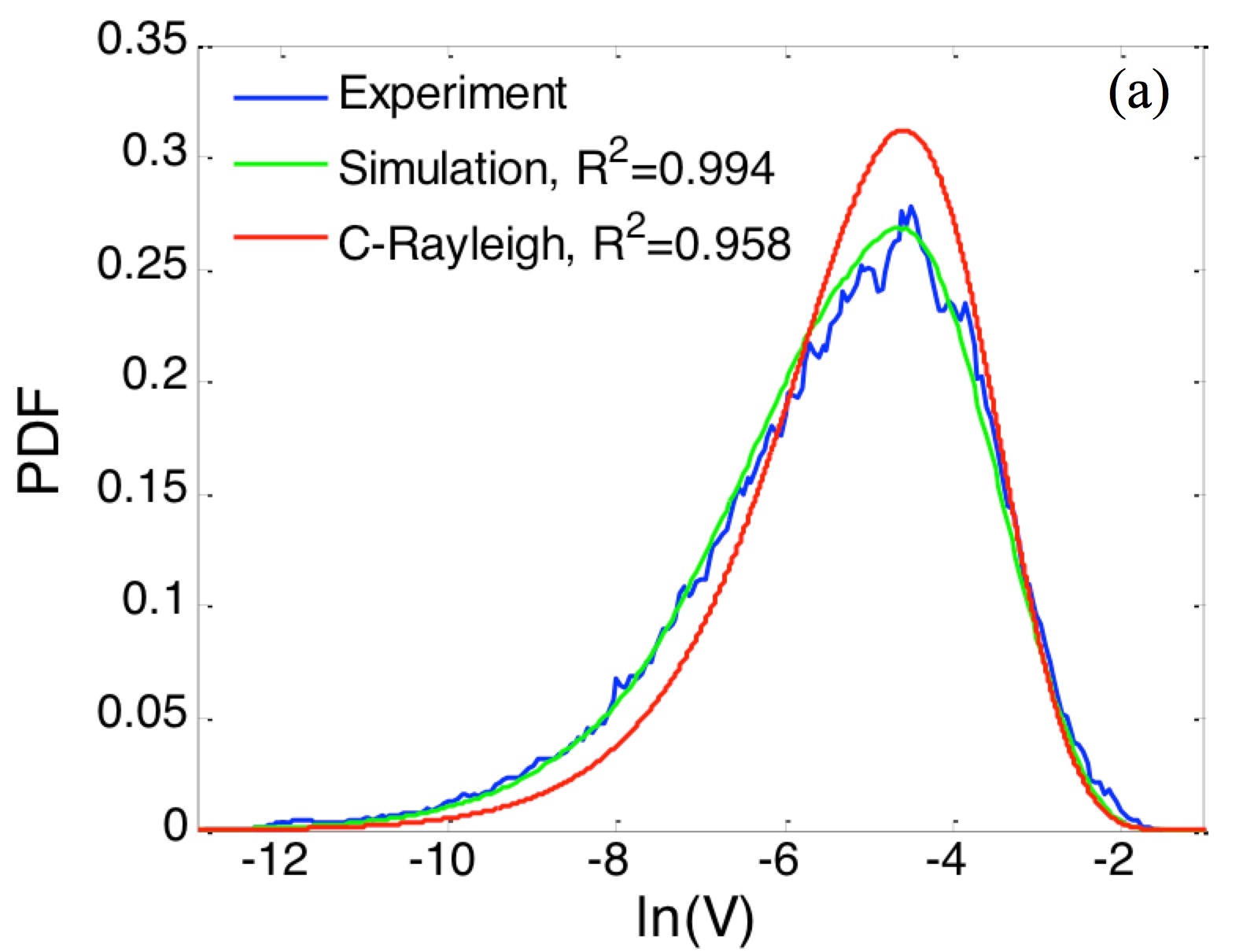}}
    \subfigure{\label{fig6b}\includegraphics[width=0.45\linewidth]{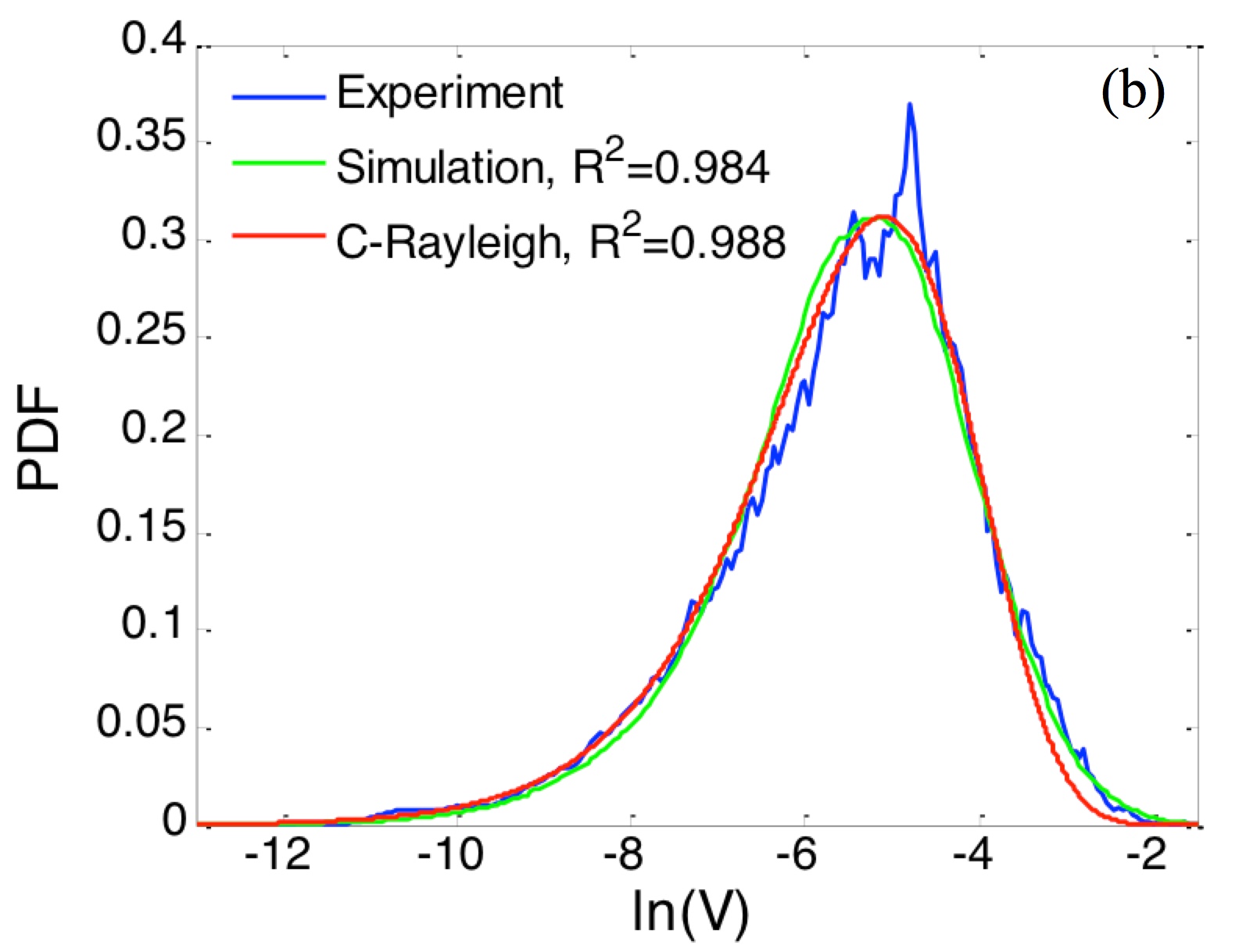}}
  \end{center}
	\caption{RCM-based \enquote{simulation} prediction (red) and Combined Rayleigh (green) fit to the second harmonic electric field between port 1 and 2 in:  (a) low loss case, $\alpha_{1}$ = 0.30, $\alpha_{2}$ = 0.78, ${P}{_{in,1f}}$ = -5 dBm, $R^2(simulation) = 0.994$, $R^2(C-Rayleigh) = 0.958$; and (b) high loss case, $\alpha_{1}$ = 2.85, $\alpha_{2}$ = 4.56, ${P}{_{in,1f}}$ = 0 dBm, $R^2(simulation) = 0.984$, $R^2(C-Rayleigh) = 0.988$. Frequency range 7 GHz - 9 GHz.}
	\label{fig6}
\end{figure}

The Combined Rayleigh distribution also has the scale parameter $\sigma$ to vary (not shown in the figure). Fig. \ref{fig6} shows the best Combined Rayleigh fits to the second harmonic induced voltage.  Since the Rayleigh fit fails in the first (linear) process in the low-loss case, the experimental second harmonic statistics deviate from the combined Rayleigh distribution as well, while both the Rayleigh and Combined Rayleigh models fit well in the high loss case, shown in Fig. \ref{fig6}. The singular RCM based \enquote{simulation} is generated for comparison. $R^2$ for this RCM-based simulation are both larger than 98\%, while $R^2 = 0.958$ in low loss and $R^2= 0.988$ in high loss for Combined Rayleigh fit.

\subsection{Double Weibull distribution of second harmonic fields}
Later a more versatile statistical model called the Double Weibull distribution was proposed to fit the mean normalized second harmonic electric field \cite{DW}. This model describes the changing statistical harmonics distribution due to the nonlinearity strength variation, i.e. the power exponent $n$ ($V_{2f} \propto V_{1f}^n$) of the nonlinear source falls below 2 as the input amplitude $V_{1f}$ increases \cite{describe}. The Combined Rayleigh fit is a special case of this model with $n=2$ for weak nonlinearity. The power exponent $n$ becomes the only fitting parameter. In the high loss case where the Rayleigh fit is valid, the fitting parameter $n$ shows a decrease from 2.11 to 1.85 as the input power increases from -5 dBm to +10 dBm. 

\begin{figure}[H]
\begin{center}    
    \subfigure{\label{fig7a}\includegraphics[width=0.45\linewidth]{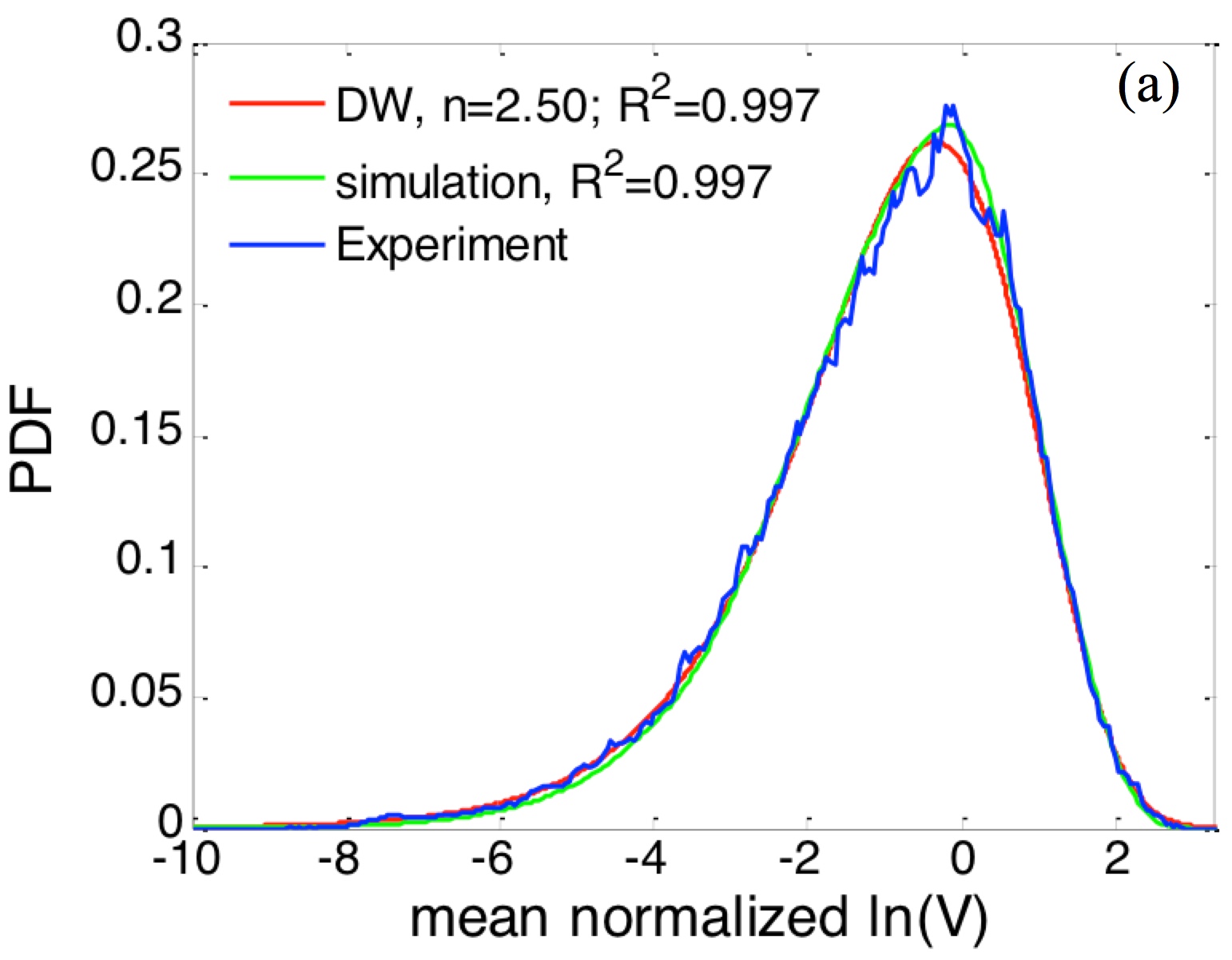}}
    \subfigure{\label{fig7b}\includegraphics[width=0.445\linewidth]{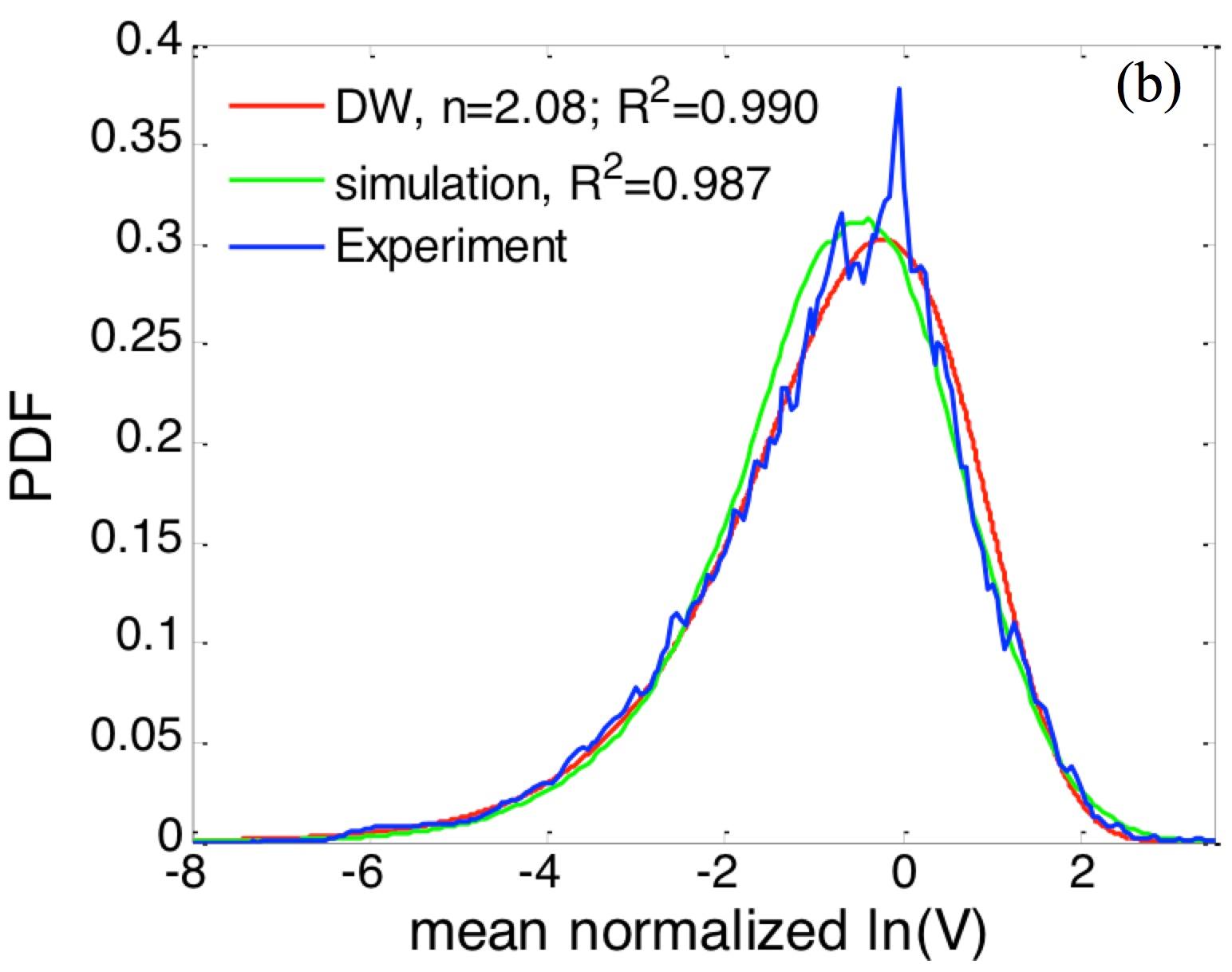}}
  \end{center}
	\caption{Double Weibull and RCM based simulation fit to the mean normalized induced 2nd harmonic voltage statistics between port 1 and 2 in:  (a) low loss case, $\alpha_{1}$ = 0.30, $\alpha_{2}$ = 0.78, ${P}{_{in,1f}}$ = -5 dBm, $R^2(simulation) = 0.997$, $R^2(D-Weibull) = 0.997$; and (b) high loss case, $\alpha_{1}$ = 2.85, $\alpha_{2}$ = 4.56, ${P}{_{in,1f}}$ = 0 dBm, $R^2(simulation)=0.987$, $R^2(D-Weibull) = 0.990$. Frequency range 7 GHz - 9 GHz of the 1/4-bowtie cavity.}
\label{fig7}
\end{figure}

However, in the low loss case where the Rayleigh fit fails, the 2nd harmonic statistics still fit well to the Double-Weibull distribution. Fig. S\ref{fig7a} shows fits to the Double-Weibull distribution with $R^2=0.997$, while the fitted $n$ is 2.50 in the low loss case. It should be noted that the nonlinear circuit is the same and has been explicitly characterized to obey a square law ($V_{2f} \propto V_{1f}^2$) in the operating regime of the experiment. A value of $n=2.5$ for second harmonic generation is therefore un-physical. Therefore the changing statistical distribution from the Combined Rayleigh distribution is not only due to changes in the nonlinearity strength but also loss of the cavity. For second harmonic statistics, stronger nonlinearity causes $n$ to decrease below 2; whereas low loss causes $n$ to increase beyond 2. In the high loss case, the fitting parameter of the Double Weibull model approaches 2, with $n=2.08$, as shown in Fig. S\ref{fig7b}. On the other hand the Random Coupling Model based approach predicts the statistics very well both in the low loss (Fig. S\ref{fig7a}, $R^2=0.997$) and high loss environment (Fig. S\ref{fig7b}, $R^2=0.987$), it demonstrates that Eq. (3) of the main text with the Random Coupling Model provides a simpler and physically reasonable explanation of the data.